\documentclass[aps,amsmath,amssymb,twocolumn,superscriptaddress,preprintnumbers,nofootinbib]{revtex4-1}
\usepackage{multirow}
\usepackage{graphicx}   

\usepackage{graphicx}

\usepackage{graphicx}
\usepackage{dcolumn}
\usepackage{bm}
\usepackage{epstopdf}
\usepackage{color}
\newcommand{\nc}{\newcommand}

\newcommand{\Eq}[1]{Eq.~(\ref{#1})}
\newcommand{\Sec}[1]{Sec.~\ref{#1}}

\newcommand{\Fig}[1]{Fig.~\ref{#1}}
\newcommand{\Tab}[1]{Table~\ref{#1}}
\newcommand{\Ref}[1]{Ref.~\cite{#1}}
\newcommand{\Refs}[1]{Refs.~\cite{#1}}
\def\lsim{\raisebox{-4pt}{$\,\stackrel{\textstyle{<}}{\sim}\,$}}
\def\gsim{\raisebox{-4pt}{$\,\stackrel{\textstyle{>}}{\sim}\,$}}
\newcommand{\be}{\begin{equation}}
\newcommand{\ee}{\end{equation}}
\newcommand{\bi}{\begin{itemize}}
\newcommand{\ei}{\end{itemize}}
\def\beq{\begin{equation}} 
\def\eeq{\end{equation}} 
\def\bea{\begin{eqnarray}} 
\def\eea{\end{eqnarray}} 
\def\ben{\begin{enumerate}} 
\def\een{\end{enumerate}}

\def\lsim{\mathrel{\raise.3ex\hbox{$<$\kern-.75em\lower1ex\hbox{$\sim$}}}} 
\def\gsim{\mathrel{\raise.3ex\hbox{$>$\kern-.75em\lower1ex\hbox{$\sim$}}}} 
\def\ifmath#1{\relax\ifmmode #1\else $#1$\fi}

\nc{\mc}{\mathcal}
\nc{\ttbar}{t\bar t}
\begin{document}

\DeclareGraphicsExtensions{.jpg,.pdf,.mps,.png,}

\title{Measuring the Bottom-Quark Forward-Central Asymmetry at the LHC}

\author{Dilani Kahawala}
\email{kahawala@physics.harvard.edu}
\affiliation{Department of Physics, Harvard University, Cambridge MA, 02138}

\author{David Krohn}
\email{dkrohn@physics.harvard.edu}
\affiliation{Department of Physics, Harvard University, Cambridge MA, 02138}

\author{Matthew J. Strassler}
\email{strassler@physics.rutgers.edu}
\affiliation{New High Energy Theory Center, Department of Physics and Astronomy, Rutgers University, 136 Frelinghuysen Rd, Piscataway, NJ 08854}

\begin{abstract}
Measurements of the top quark forward-backward asymmetry performed at the Tevatron suggest that new-physics may be playing  a role in $t\bar t$ production.  To better understand the source of the asymmetry, recent proposals have called for a measurement of the bottom and charm forward-backward asymmetries at the Tevatron, using jets with embedded muons.  Here we propose a corresponding measurement of the bottom quark forward-central asymmetry designed to look for similar effects in the $b$-sector at ATLAS and CMS.  We construct a set of 
cuts designed to enhance sensitivity to this asymmetry, and test our analysis on  a toy axigluon model representative of those used to explain the top asymmetry. We find that if the relevant new-physics couplings to the bottom quark are similar to those of the top, then the effects should be visible at the $2\sigma$ level in less than $10~{\rm fb}^{-1}$ of $7~{\rm TeV}$ LHC data.  Such a measurement would be of general importance, and would provide valuable model-building input, serving to restrict the set of models put forward to explain the Tevatron $t\bar t$ anomaly.  However, a relatively low trigger threshold on non-isolated muons inside hard jets must be maintained to allow for this measurement.   

\end{abstract}

\maketitle

\section{Introduction}

Analyses by the CDF \cite{CDF-leptonic:2011,*Aaltonen:2011kc,*Aaltonen:2008hc} and D0 \cite{DO:2007qb,*Abazov:2011rq} collaborations suggest that the top forward-backward asymmetry is much larger than predicted by the Standard Model (SM). This asymmetry, which essentially measures the extent to which the top
in $t\bar t$ production has a preference to be aligned with the initial state quark (rather than anti-quark) is only a few
percent within the SM, yet has been measured to be ${\cal O} (10\%) $ in inclusive $t\bar t$ samples, and even ${\cal O}(40\%)$
at high $m_{t\bar t}$.  The latest results from CDF further demonstrate that the large asymmetry manifests itself at the $3\sigma$ level in both the semi- and fully-leptonic $t\bar t$ decay channels,  making naive systematic/statistical  effects a less likely explanation for the effect.  While recent results from D0~\cite{Abazov:2011rq} disagree somewhat with those of CDF in several areas (most importantly in the behavior of $A_{FB}$ at large invariant mass), and the LHC has yet to observe 
evidence for any similar new-physics effects~\cite{CMS-PAS-TOP-11-014,*ATLAS-CONF-2011-106}, this asymmetry remains one of the most compelling experimental anomalies.

Indeed, as the top asymmetry continues to resist a more conventional explanation, many models of new-physics have been put forth to explain the anomaly~\cite{Bai:2011ed,delAguila:2011yd, *Giudice:2011ak, *Djouadi:2011aj, *Barcelo:2011fw, *Chen:2011mg, *Nelson:2011us, *Wang:2011ta, *Degrande:2011rt, *Jung:2011ua, *Zhu:2011ww, *Fox:2011qd, *Jung:2011ue, *Shu:2011au, *Rajaraman:2011rw, *AguilarSaavedra:2011zy, *Buckley:2011vc, *Jung:2011zv, *Casagrande:2011hf, *Ligeti:2011vt, *Grojean:2011vu, *Blum:2011up, *Foot:2011xu, *Barreto:2011au, *Zerwekh:2011wf, *Isidori:2011dp,*Patel:2011eh, *Grinstein:2011yv, *Barger:2011ih,*Shelton:2011hq,*Delaunay:2011vv,*Cheung:2011qa,*Cui:2011xy,*Tavares:2011zg,*AguilarSaavedra:2011ci}, see \Ref{Kamenik:2011wt} for a review. 
These models typically introduce new heavy intermediate particles to generate the top asymmetry via interference with the SM, and differ principally in (1) whether they are  $s$, $t$,  or $u$-channel, (2) in the spin/color of the new degrees of freedom, and (3) in their couplings to the first and third generation quarks.
While it is true that many of these models seem to be in tension with other measurements (especially the $t\bar t$ differential cross section and searches for same-sign tops at the LHC~\cite{Collaboration:2011dk}), such considerations can be subtle~\cite{Gresham:2011pa,Gresham:2011fx} and various models may still be able to reproduce the $t\bar t$ asymmetry while maintaining consistency  with other measured properties of the top.  Thus, to make progress in understanding the origin of the top asymmetry it is helpful to keep an open mind toward new models and subject them to further experimental scrutiny. Many analyses have already been proposed with this aim, including an LHC measurement of the top-quark forward-central asymmetry~\cite{Wang:2010du,*Hewett:2011wz,*Xiao:2011kp,*Bhattacherjee:2011nr,*AguilarSaavedra:2011hz}, studies making use of the polarization/spin-correlation in the $t\bar t$ system~\cite{Berger:2011hn,*Godbole:2010kr,*Cao:2010nw,*Jung:2010yn,*Choudhury:2010cd,*Bernreuther:2010ny,*Melnikov:2009dn,*Krohn:2011tw,*Bai:2011uk}, and more specialized analysis designed to look for the signatures of  particular models~\cite{Gresham:2011dg,*Craig:2011an,*Cao:2011ew,*Alvarez:2011hi,*Berger:2011ua}.   

Since the flavor structure of the various models differs widely, it is important to measure similar asymmetries for other quarks.  To wit, if the asymmetry in $t\bar t$ is indeed due to the effects of new physics, then one must ask if these only apply to the top-sector, or  if they affect the entire third generation of fermions.  Models of $t/u$-channel physics, for example, tend to affect only the right-handed top (or, in some cases, the entire up-type sector), using a flavor off-diagonal interaction to couple it to a first generation $u$ or $d$.  In contrast, the simplest axi-gluon models couple new-physics with opposite signs to the left and right handed tops, and so necessarily include new couplings to the bottom sector.  

It has  been pointed out recently~\cite{Strassler:2011vr,Bai:2011ed} that the data sets at the Tevatron are large enough to allow interesting measurements of the forward-backward asymmetries of both bottom and charm quarks in the same kinematic regime in which the top asymmetry is observed by CDF.  This can be done with a dijet sample, using the charge asymmetries of muons embedded in high-$p_T$ jets.  The muon charge asymmetry is correlated with the charge asymmetries of the main sources of muons, namely $c$ and $b$ quark decays.  The forward-backward asymmetry prior to heavy-flavor tagging is dominated \cite{Strassler:2011vr} by a combination of the $c\bar c$ and $b\bar b$ asymmetry, and separating bottom from charm can be done using heavy-flavor tagging and kinematics.   This analysis could help discern the different signatures of the various classes of models, especially when used in concert with some of the other tools referenced above.

In the current paper we consider a similar measurement at the LHC.  We will limit ourselves to the bottom
quark asymmetry.  This is because the dilution of the asymmetries from symmetric backgrounds is much larger at the
LHC than at the Tevatron, making charm asymmetries extremely difficult to detect.\footnote{With an asymmetry comparable to that seen in $t\bar t$ samples, the observed raw asymmetry prior to heavy-flavor tagging would be of order 2--3\% at the Tevatron \cite{Strassler:2011vr}.  But at the LHC it would be a factor of 10 smaller,  presumably too low for beating systematic errors.  Only with heavy-flavor tagging can the observed asymmetries at the LHC reach the percent level and above, but tagging removes most of the charm sample, leaving sensitivity only to $b\bar b$ physics.}  

 Clearly, in contrast to the Tevatron and its beams of opposite charge, one cannot as simply measure a forward-backward asymmetry at a parity-symmetric collider like the LHC, whose beams are both of protons.  But one can instead make use of the fact that in quark-antiquark collisions in a proton-proton machine,  the motion of the parton center-of-mass frame relative to the lab frame is correlated with the direction of the incoming quark.   Thus one may define a  ``forward-central'' asymmetry, looking at whether $b$ quarks tend to be at higher rapidity $|y|$ on average than $\bar b$ antiquarks.  A corresponding observable is employed by most  analyses which propose to measure the $t\bar t$ asymmetry at the LHC, e.g. \Ref{Wang:2010du,*Hewett:2011wz,*Xiao:2011kp,*Bhattacherjee:2011nr}.

As in \cite{Strassler:2011vr,Bai:2011ed},
following \cite{Sehgal:1987wi}, we will use the charge of a muon
embedded in a jet to determine whether the parent of the jet is more
likely to be a $b$ or
a $\bar b$.  The muon also provides us with an object for
triggering.  We will also use $b$ tagging and kinematic cuts to reduce
backgrounds.  As we will see, the measurement is difficult, although
potentially feasible.  For an underlying asymmetry of the size needed
to explain the CDF $t\bar t$ anomaly, we are only able to obtain an
observable asymmetry of order 2\% or less, which is several times
smaller than the corresponding forward-backward asymmetry at the Tevatron, 
yet we expect  such an asymmetry to be visible above the  $2\sigma$ 
level in   10 ${\rm fb}^{-1}$.  We note that an asymmetry of the size expected in the Standard Model should be
unobservable for the foreseeable future.  The Tevatron data set may
well be an easier place to make the measurement.  However, we have
certainly not exhausted all the options for improving the
signal-to-background ratio at the LHC, and we feel our result should
be viewed as encouraging, though in need of improvement by more
sophisticated means.

This paper is structured as follows.  In \Sec{sec:observable} we define the forward-central asymmetry carefully. Later, in \Sec{sec:lhcanalysis}, we will describe a set of cuts designed to optimize the discriminating power of  this quantity in the kinematic region relevant for the top asymmetry.  {\it We emphasize that the signal region lies precariously close to the trigger thresholds, and some consideration of trigger strategy must be made in the near future if one is to ensure that the data relevant for this measurement is actually recorded.}   Introducing a signal comparable to that observed at the Tevatron (using a conservative axigluon toy model) we will show that a 7~TeV LHC  can resolve an asymmetry in $b$-quark  production at more than $2\sigma$ in $10~{\rm fb}^{-1}$.  While this level of statistical significance is not sufficient to claim the discovery  of new phenomena,  it would provide helpful model-building input, allowing ATLAS and CMS data to restrict the set of models which have been put forward to explain the $t\bar t$ anomaly.   We comment on  various experimentally relevant issues and the prospects for an LHCb measurement of the asymmetry in  \Sec{sec:discussion}.  We conclude in \Sec{sec:conclude}.

\section{Observable}
\label{sec:observable}
There are two natural forward-backward asymmetries to consider
at a proton-antiproton machine, as applied to $b\bar b$ production.  
The first is to define
forward and backward in the lab frame
\begin{equation}
\label{eq:fb}
A^{b\bar b,{\rm lab}}_{FB}=\frac{N (q y>0)-N (q y<0)} {N (q y>0)+N (q y<0)},
\end{equation}
where $q=1$ $(-1)$ for the $\bar b$ ($b$) that generates the jet containing the observed muon,
and $y$ is its rapidity\footnote{One could construct a similar analysis employing
 pseudo-rapidity ($\eta$) instead of rapidity($y$).  As we are considering objects for which 
 $m\ll p_T$, the results obtained would be largely the same.}.
Here $y\to+\infty$ ($-\infty$) is the direction of motion of the proton (antiproton).
But the event-by-event boost of the hard-scattering system tends to wash
out this variable, so it is better to consider
forward and backward defined in the hard-scattering rest frame
\begin{equation}
\label{eq:fb}
A^{b\bar b}_{FB}=\frac{N (q \Delta y>0)-N (q \Delta y<0)} {N (q \Delta y>0)+N (q \Delta y<0)},
\end{equation}
where $\Delta y$ is the signed rapidity difference between the $b$ and the $\bar b$ ({\it i.e.},
the rapidity difference between the two jets signed by the muon charge.)

At a proton-proton machine such as the LHC, symmetric under $y\to -y$, these forward-backward asymmetries will necessarily be zero.
Instead we must turn to a forward-central asymmetry, which we define as:
\begin{equation}
\label{eq:fc}
A^{b\bar b}_{FC}=\frac{N (q\Delta| y|>0)-N (q\Delta| y|<0)}{N (q\Delta| y|>0)+N (q\Delta| y|<0)},
\end{equation}
where now \(\Delta| y| = |y({ b})|-|y({\bar b})|\) is defined as the
rapidity difference between the rapidity of the $b$ and the $\bar
b$.\footnote{We note that a differential distribution of the
asymmetry, e.g. $dA^{b\bar b}_{FC}/d\Delta |y|$ or $dA^{b\bar
b}_{FC}/dy_{jj}$ (see Eq.~\ref{eq:avg_rap}) may provide an even more
powerful discriminant, although for simplicity we will not consider
these here.}  

If we consider a $q\bar q$-initiated scattering
process in a $pp$ system, the direction of the boost of the
hard-scattering system along the beam direction will tend to reflect
the direction of the initial state quark.  This effect is illustrated
in Fig.~\ref{fig:rapcut}.

It is instructive to consider the behavior of Eqs.~(\ref{eq:fb}) and (\ref{eq:fc})
under various reflection symmetries to determine their susceptibility to shifts from experimental errors.
The forward-backward asymmetry measured at the Tevatron, for instance, flips sign ($A^{b\bar b}_{FB}\rightarrow -A^{b\bar b}_{FB}$) if either
$y\to -y$ or $q\to
-q$.  This tells us if there were no asymmetry to begin with, one would not be induced via
a distortion in the efficiency to measure one charge over the other as long as both sides of the detector saw the same distortion.
That is, the only way
to find a spuriously non-zero value in $A^{b\bar b}_{FB}$ would be to introduce a distortion in the charge efficiencies which was not invariant under $y\rightarrow -y$.
The situation at the LHC is more subtle as $A^{b\bar b}_{FC}\rightarrow +A^{b\bar b}_{FC}$ under $y\rightarrow -y$, 
but one still has $A^{b\bar b}_{FC}\rightarrow -A^{b\bar b}_{FC}$ under $q\to -q$, which tells us that to the extent the
detection efficiencies for muons and antimuons are equal, at any given
rapidity, it is still the case that no asymmetry can
be generated if none exists.   
We therefore emphasize that while every effort should be made to correct for detector and trigger effects to
obtain a reliable measurement,
the forward-central asymmetry of \Eq{eq:fc} is fairly robust against systematic shifts from rapidity-dependent efficiencies.

 Of course, the asymmetry (or limit on an asymmetry) observed in data
must be converted into an asymmetry (or limit) in the underlying
$q\bar q\to b\bar b$ process.  This translation will require careful
modeling of the muon efficiency as a function of $y$.  But this last
is also true for the Tevatron measurement, which involves an integral
over $y$, so there too one must account for the $y$-dependent detection
efficiencies.

\section{LHC Analysis}
\label{sec:lhcanalysis}
\begin{table}
\begin{ruledtabular} 
\begin{tabular}{cl}
&{\bf Jets}: We require at least two jets with $|y(j)|$\\
 & \ \ $<2.4$,  and further demand $p_T(j_1)>150~{\rm GeV}$\\
 &\ \ and  $p_T(j_2)>100~{\rm GeV}$.\\ 
Selection& {\bf Muon:} There must be a $\mu$ close to $j_1$ or $j_2$ \\
cuts&\ \ satisfying  $\Delta R(j,\mu)<1$, $p_T(\mu)>25~{\rm GeV}$, \\
&\ \  and $|y(\mu)| < 2.4$\\
&{\bf Flavor tag:} Finally, we require that the jet \\
& \ \ {\it without} the nearby muon is $b$-tagged.\\ 
 \multirow{2}{*}{Forward  cut} & \multirow{2}{*}{$|\frac{y(j_1)+y(j_2)}{2}|>0.5$}\\
\\ 
\multirow{2}{*}{Mass  cut} & \multirow{2}{*}{$m({j_1+j_2})>450~{\rm GeV}$}\\
\\
\end{tabular}
\end{ruledtabular}
\vspace{0.3cm}
\caption{Cuts used to select events and to increase the signal size.  We denote the $i$-th hardest jet as $j_i$.  The effect
of the cuts can be seen in  \Tab{tab:sig}.\label{tab:cuts}}
\vspace{0.3cm}
\end{table}

As in~\cite{Strassler:2011vr,Bai:2011ed}, our strategy is to consider
dijet events in which one of the two leading jets
has an embedded muon, and to use the muon's charge as
an approximate surrogate for the charge of the parent
$b$ quark~\cite{Sehgal:1987wi}.
The resulting forward-central asymmetry in
charged non-isolated muons is diluted by many effects, to be discussed
below, but its value does correlate with the  
forward-central asymmetry in $q\bar q \to b\bar b$ events we wish to measure.

Let us first define our event sample. We will assume
that a trigger exists that can easily accommodate a single non-isolated
muon of 25 GeV within an event with at least
one jet of 150 GeV and $H_T$ of at least 250 GeV.\footnote{We note that the
cross section for events passing this particular trigger stream is roughly 1 nb 
at a 7 TeV LHC, and so could be easily accommodated in ${\cal L}\sim 10^{33}{\rm cm}^{-2}{\rm s}^{-1}$ running.}
We will
see this accords with the requirements of the measurement.
Significantly higher thresholds might put the measurement
out of reach.  Within this sample we demand the jets be di-jet-like\footnote{Events that differ 
strongly from dijet structure ---
for instance, those in which the two leading jets are not fairly back-to-back or have
MET that does not point roughly in the direction of one of the leading jets --- should
be vetoed.  However these vetoes should be loose and chosen 
with some care to avoid making theoretical calculations of backgrounds
unstable.}
and veto events with
an isolated lepton.

To put ourselves in the same mass region as is probed
by the measurements of the $t\bar t$ asymmetry, we will focus
on dijet events where the hardest jet's $p_T$ is greater
than 150 GeV and the second hardest's is greater than
100 GeV. As we will later demand a muon in one jet and
a b-tag on the other, we require that both jets lie within
$|y| < 2.4$ so they are within the tracking system. Later we
will see that an additional cut requiring $m_{jj}> 450$ GeV
will help us to further increase the signal to background
ratio, although it will not help increase the statistical
significance of the results. 

\begin{figure}
\includegraphics[scale=1.0]{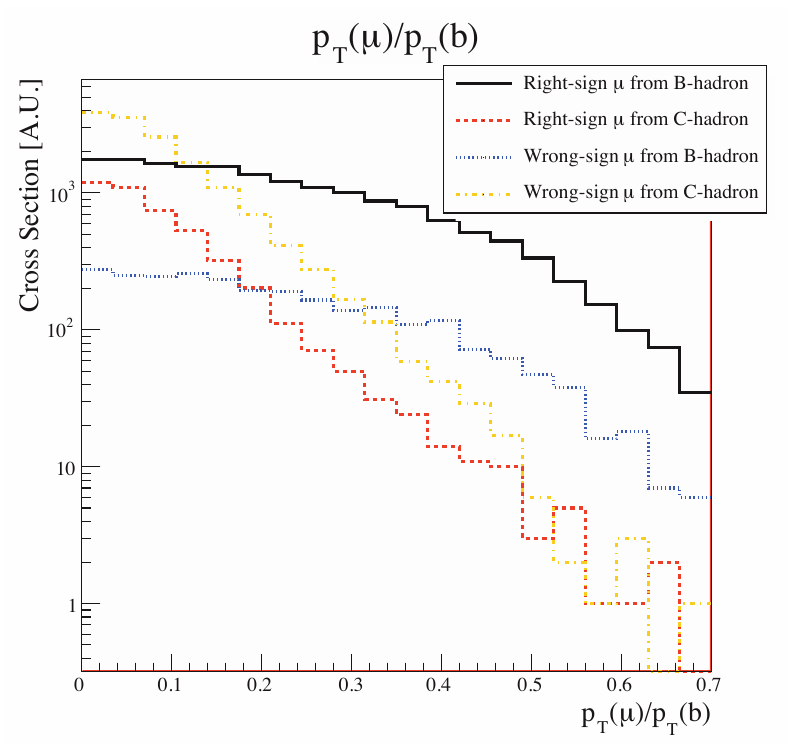}
\caption{The fractional $p_T$ carried by a muon produced from the hadrons of a $b$-jet with $p_T>150~{\rm GeV}$ where we have distinguished right-sign muons ($b/\bar b\leftrightarrow \mu^-/\mu^+$) from wrong-sign muons($b/\bar b\leftrightarrow \mu^+/\mu^-$).  \label{fig:muz}}
\vspace{0.3cm}
\end{figure}
\begin{figure}
\begin{center}
\includegraphics[scale=1.0]{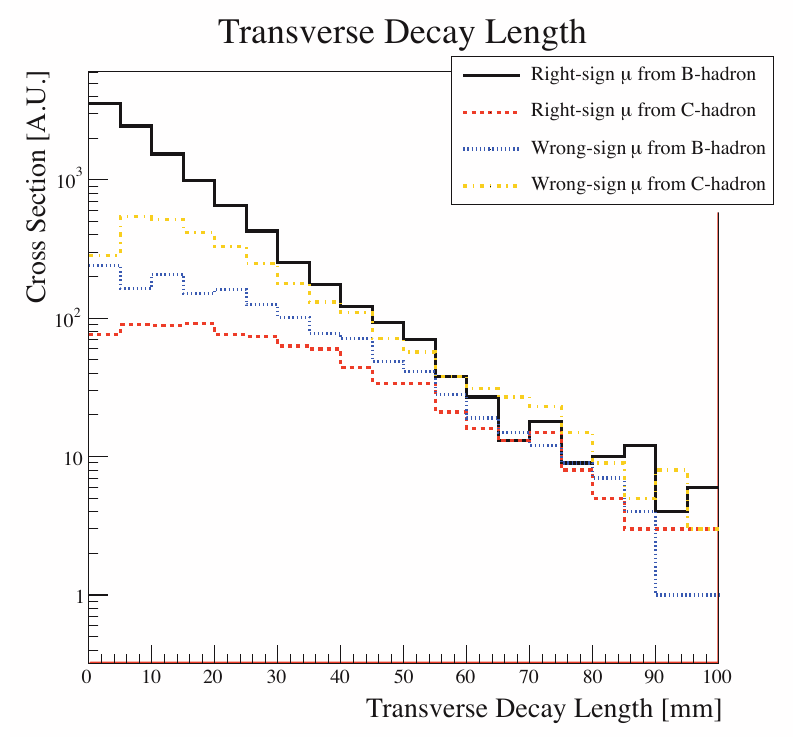}
\end{center}
\caption{The transverse decay length of the secondary vertex which gives rise to the muon.  Here we have required $b$-jets with $p_T>150~{\rm GeV}$ and $p_T(\mu)>25~{\rm GeV}$.\label{fig:tdl}}
\end{figure}

The charge of the muon does not provide a fully reliable measure of
the initiating quark's charge, due to a combination of effects. The
largest of these, at least when the muon carries a low fraction of the
jet's $p_T$, comes from the decay of a $b$-quark into a $c$, whose
subsequent decay produces a wrong-signed muon. Another cause of wrong
signed events is from neutral $B$-meson oscillations which switch
constituent $b$'s to $\bar b$'s, and vice-versa. To reduce the
concentration of wrong-signed events, it is helpful, for jets with
$p_T$ of order 150 GeV, to take the muon inside the jet to have $p_T$
above 25 GeV. This is illustrated in \Fig{fig:muz}, where we show the
relative fraction of the jet $p_T$ carried by the muon for $b$-jets
with $p_T>$ 150 GeV. As emphasized above, we are assuming here that 25
GeV is compatible with an available trigger pathway.  This assumption
appears to be correct at current luminosities but may not necessarily
remain so throughout the 2011-2012 run.

Only a fraction of the selected events will contain a $b\bar b$
final state, with other contributions from $c\bar c$ pairs, gluons
that split to heavy flavor, and events with a single $b$ or
$c$. We significantly reduce the unwanted contributions
by demanding that among the two hardest jets, the jet that does
not contain the muon is $b$-tagged.\footnote{Tagging the jet
containing the muon would not assist, since the relatively hard non-isolated muon is already
indicative of a $b$ jet. However, see the discussion section below.}
This is our ``initial selection''. Already, as we will see,
there is sensitivity to an asymmetry comparable to that seen in the $t\bar t$ system.
But the initial sample is not very pure, and
the observed asymmetry would be very small, less than one percent. To increase
the size of the observed asymmetry, with limited
effect on the sensivity, we impose additional cuts.

\begin{figure*}
\includegraphics{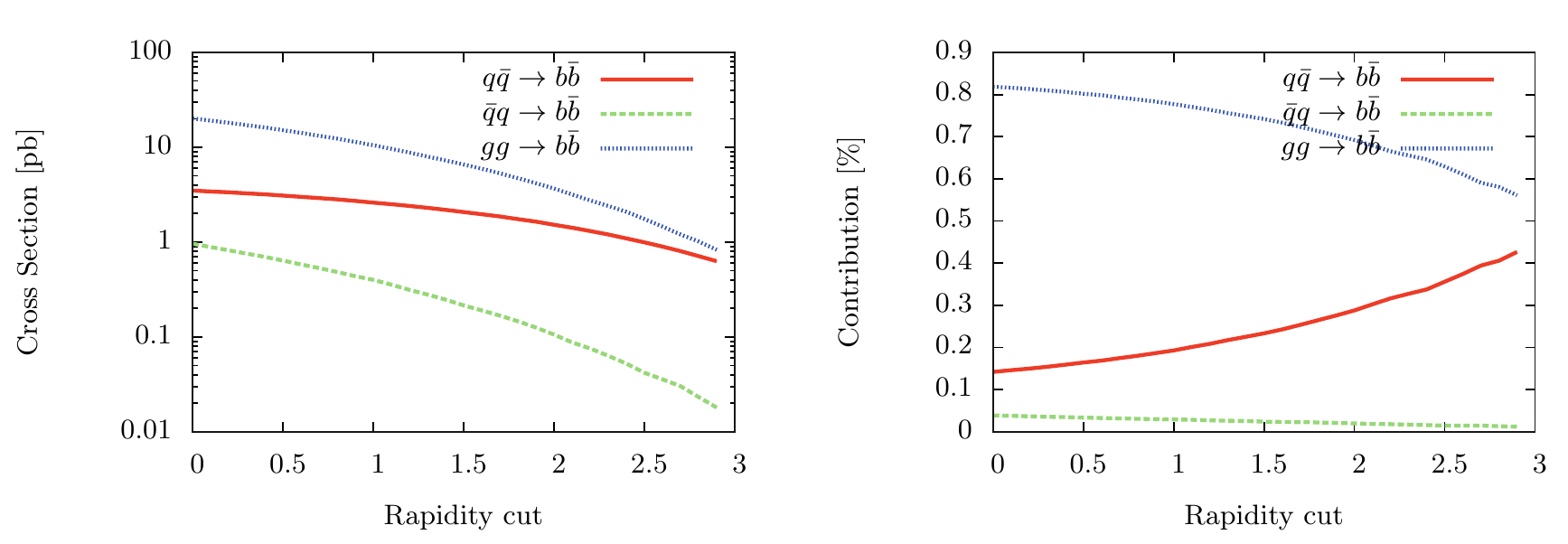}
\caption{Left: The cross-section for $b\bar b$ production as a function of the minimum  rapidity cut on $y_{jj}=[y(j_1)+y(j_2)]/2$ for various initial states subject to the $150~{\rm GeV}$ $p_T$ cut
used in the analysis.  For the $q\bar q$ initial state the quark moves in the direction of the center-of-mass as measured
in the laboratory, while for the $\bar q q$ state the anti-quark moves in this direction.  
Right: The relative contribution from each initial state to the $b\bar b$ sample as a function of the minimum rapidity cut. \label{fig:rapcut}}
\vspace{0.7cm}
\end{figure*}

In our initial selection sample, the symmetric process $gg \to b\bar b$
dominates over the process we hope to probe: $q\bar q \to b\bar b$.
We can attempt to mitigate this dilution by using the
fact that, at fixed $H_T$, the partonic center-of-mass frame in $gg$
events tends to be at lower $|y|$ than in $q\bar q$ events.
We define the average rapidity of the dijet system to be
\be
\label{eq:avg_rap}
y_{jj}=\frac{y(j_1)+y(j_2)}{2}
\ee
and apply a cut $|y_{jj} | > y_{\rm min}=0.5$ requiring that the system
be relatively forward. The effects of such a cut can be
seen in \Fig{fig:rapcut} where we show how such a cut increases
the relative contribution of `right-orientation' $q\bar q\to b\bar b$ events --- ones where the
quark is moving in the same center-of-mass direction as
the lab-frame $b\bar b$ system --- while decreasing both the symmetric
($gg\to b\bar b$) and `wrong-orientation' $\bar q q\to b\bar b$ (where the anti-quark moves in the
same direction as the scattering system) backgrounds.
For the sake of simplicity we will only employ a sharp cut
on $|y_{jj}|$, but a more powerful analysis could make use of a
continuous discriminant to achieve a higher significance.

Furthermore, as we wish to measure the extent to
which any new physics giving rise to the top asymmetry
might be affecting the bottom sector, the best place
to look for it is probably in the same kinematic region where the
top anomaly is observed by CDF, namely at $t\bar t$ invariant mass above 450
GeV. We will therefore focus on events with a similar requirement
that $m_{jj} > 450$ GeV.  (Recent results from D0 might lead one to
rethink this step, but in any case, as we will see, this cut is useful but not essential.)
This also reduces backgrounds more than signal.

Our cuts are summarized in Table I. To quantify our
results we introduce an asymmetric $b\bar b$ signal that is comparable, in the appropriate invariant-mass range, to that seen in $t\bar t$ at the
Tevatron.  We do this in a way that we believe is conservative.  In particular, we use a toy model similar to the minimal two
site axigluon model described in Ref. \cite{Bai:2011ed}. We choose
the axigluon mass to be \(M_{G^{'}}=1041 \) GeV and the width to be  \(\Gamma({G^{'}})=200 \) GeV; 
we set its vector-like coupling to
zero, and set its axial coupling equal in strength
to the QCD coupling.\footnote{The sign of the coupling of the axial gluon to the
first generation is opposite to that of the coupling to the third
generation.} Using these parameters we find
a rest-frame parton-level asymmetry in acceptable agreement with that observed in the top system at both
CDF and D0: 8\% for invariant mass less than 450 GeV and 30\%
for invariant masses above 450 GeV.  To avoid assumptions about what occurs above the reach of the Tevatron,
we only consider events (the significant majority, in any case) with
$m_{jj}< 800$ GeV.  This is well below the axigluon pole, which avoids sensitivity to the model-dependent resonance structure
of the axigluon.  Note that although our toy model may itself be excluded by 2011 LHC data, we are only using aspects of the model that are largely model-independent.  For instance, a different toy model with a larger axigluon mass and coupling would give a similar signal to the choice made here for $m_{jj}< 800$ GeV.  Any additional contributions to the asymmetry above 800 GeV will only make new-physics effects more prominent.

Before we proceed we describe our Monte-Carlo analysis tools: we use \texttt{Madgraph v.4.4.60}  \cite{Alwall:2007st} to generate our parton-level $gg\rightarrow b\bar b$ and $q\bar q\rightarrow b\bar b$ samples, which are then showered in \texttt{Pythia v.6.4.24} \cite{Sjostrand:2006za}. Both of these generators are used with their default PDF distributions: CTEQ~\cite{Pumplin:2002vw} 6L1 and 5L, respectively.
Our  calculations ignore the next to leading-order
asymmetry in $b$-production present in the SM, although we expect\footnote{From \Ref{Kuhn:1998jr} we estimate that the asymmetry in $b$ production 
at the Tevatron is \(\mathcal{O}\)(6\%) at high energies, to be compared to the
${\cal O}(30\%)$ one might expect from new-physics effects. } this to be subdominant compared to the new-physics effects we consider, and would at most 
yield a small linear shift in our results. 
The other processes which constitute our background\footnote{The backgrounds for our study, aside from the symmetric $gg\rightarrow b\bar b$ process which we generate with Madgraph, include process with light-jets, processes with charm jets, gluon-splitting processes (e.g. $g\rightarrow b\bar b$) and flavor-excitation processes ($bx\rightarrow bx$).} are  generated and showered in Pythia. After generation, all events are clustered into jets between $-5<\eta<5$ by \texttt{Fastjet v.2.4.2} \cite{Cacciari:Fastjet,Cacciari:2005hq} using the anti-\(k_T\) algorithm \cite{Cacciari:2008gp} 
with \(R=0.7\). We will assume a \(b\)-tagging efficiency of $50\%$, a \(c\)-fake rate of $10\%$, and a light-jet fake rate of $0.3\%$  \cite{Aad:2009wy}.\footnote{More precisely, we scan inside a jet for a $b$- or $c$-containing hadron, and tag with the above efficiencies if one is found.}  

It is instructive to study the composition of processes contributing to the non $q\bar q/gg\rightarrow b\bar b$ background (i.e., what goes into the ``other background'' row in  \Tab{tab:sig}).  The largest
component of these events comes from the production of events with real $b$-quarks, the 
rate for which we find to be roughly comparable to that of $gg\rightarrow b\bar b$ processes.  Of these, approximately $60\%$ come from flavor excitation processes and the rest come from 
gluon splitting in the parton shower.  The next largest contribution comes from processes in which 
a $c$ fakes a $b$.  The rate for this process, before accounting for the efficiency of a 
$c$ to fake a $b$, is roughly twice the rate for $gg\rightarrow b\bar b$ (again, before tagging efficiencies).  Finally, the raw rate for processes yielding a muon in one jet
but no $c/b$-hadrons in the other is usually  ten to twenty times the rate for $gg\rightarrow b\bar b$ production, 
but once one applies the $0.3\%$ efficiency for light QCD to fake a $b$-jet, this process makes the 
smallest contribution of those considered.  Now, since many of our jets are at high $p_T$ and high $y$, this may be a low estimate on the light QCD mistag rate.  However, the rate is so small that our result is not very sensitive to the mistag rate, which could be as large as a few percent without having a qualitative impact on our conclusions.

\begin{table}
\begin{center}
\begin{ruledtabular} 
\begin{tabular}{cccc}
& Selection & $y_{jj}>1/2$&  $m_{jj}>450 $    \\
 \hline 
  $\sigma_{q\bar q \rightarrow b \bar b}$ (pb) & 1.1 & 0.9 & 0.3\\
  $\sigma_{\bar q q \rightarrow b \bar b}$ (pb) & 0.3 & 0.1 & 0.0\\
  $\sigma_{gg \rightarrow b \bar b}$ (pb) & 7.1 & 4.0 & 0.9\\
other background & 10.0 & 5.7 & 1.6 \\
 \hline
$\sigma_{\rm total}({\rm pb})   $ & $18.6$ & $10.7$ & $2.7$\\
$A^{b\bar b}_{FC}(\%)$ & $0.6$ & $0.9$ & $1.6$\\
significance ($\sigma$)  & $2.5$ & $2.8$ & $2.6$
\end{tabular}
\end{ruledtabular}
\caption{\label{tab:sig}The rates of various contributing processes, the forward-central asymmetry and and the statistical significance after selection, rapidity and invariant mass cuts (the cuts are presented in \Tab{tab:cuts}). We denote by $q\bar q \rightarrow b \bar b$ the `right-orientation' $q\bar q$ intial state, and by $\bar q q \rightarrow b \bar b$ the `wrong-orientation' state.  Our `other background' contribution includes processes of flavor excitation and gluon splitting, as well as fake $b$'s from charm and light flavor.  The results account for a tagging efficiency of 50\%/10\%/0.3\% for $b/c$/light-flavor jets. The significance is measured as \(1/\sqrt{N}\) assuming ${\cal L}=10~{\rm fb}^{-1}$.}
\end{center}
\end{table}

Our results showing the behavior of $A_{FC}^{b\bar b}$ as various cuts are imposed
are presented in \Tab{tab:sig}.  Beginning with the initial selection
cuts we first apply a cut on $y_{jj}$ which increases the observed asymmetry 
by $\sim 50\%$, and next apply a cut on $m_{jj}$ to increase the
observed asymmetry by another $\sim 80\%$.  While both of these cuts have a
minimal impact on the statistical significance of the measured asymmetry, the
increase in its absolute size is comforting as it reduces the impact
of systematic errors.  A sensitivity of more than $2\sigma$ is possible with about 10 fb$^{-1}$ at 7 TeV.

\section{Discussion}
\label{sec:discussion}

Let us first make a brief theoretical comment before turning to the more serious experimental issues. In presenting the estimates of the previous section we have aimed to remain relatively conservative.  Our toy model yields a somewhat small asymmetry compared 
to the CDF results, and if new physics is indeed present it may generate larger effects than we considered and would therefore manifest itself sooner.  We have not accounted properly for K factors, but they tend
to be larger than 1 for QCD di-jet processes.  Accounting for them is unlikely to change the signal-to-background ratio very much though the statistical significance we found may slightly improve (though admittedly the improvement is likely to be cancelled by experimental inefficiencies.)  We have no reason to expect unusually large K factors given that we have not introduced restrictive cuts on phase space.  Also we should emphasize that a change in the relative rates of the different contributing processes will not induce a new source of asymmetry.

We believe that dominant sources of theoretical uncertainty affecting the analysis we propose are 
probably the uncertainties on (a) the values of the NLO K-factors, which will affect all of the production rates, (b) the gluon, $c$, and $b$ PDFs, which are important in the backgrounds, and (c) the process of gluon splitting to $c\bar c$ and $b\bar b$ within a jet, which is also important for the backgrounds.  The uncertainty on the Standard Model prediction for $A_{FC}^{b\bar b}$ is likely unimportant.  We further note that there are a number of data-driven handles that might be useful for determining backgrounds, including observables such as (a) the probability for a jet to contain two $b$-tags or multiple muons (both same- and opposite-sign), (b) the embedded muon $p_T$ and $k_T$ spectra (where $k_T$ is measured with respect to the jet axis), and (c) tracking/vertexing information on $b$/$c$ hadrons within jets.

A serious concern that we cannot address here involves the trigger.  It is not clear to us that the cuts required for the analysis are compatible with the triggers that will be used in accumulating $10~{\rm fb}^{-1}$ of data.   Again, the ingredients of the analysis are simple: a dijet event with the leading (sub-leading) jet carrying 150 (100) GeV of $p_T$,  one of the two jets containing a muon with $25$ GeV of $p_T$ and the other $b$-tagged.  A non-isolated-muon-plus-$H_T$ trigger might be a suitable pathway, perhaps supplemented at higher trigger levels by requiring at least one of the two jets to contain displaced tracks.  Requiring the muon track in particular to be slightly displaced is another possibility, but comes at the high cost of reduced statistics and possible challenges for trigger-acceptance determination.  We must leave these important details to our experimental colleagues.

There is potentially additional room for the experiments to improve upon the analysis we have presented. The most obvious step would be to include electronic decays in addition
to the muons used here, but since electrons come with a higher trigger threshold it is not clear this would add much sensitivity. 
Another potential source of improvement\footnote{We thank Gustaaf Brooijmans for bringing this to our attention.}  
could come from using the  displacement distance of displaced vertices to reduce the
dilution of the underlying asymmetry from neutral $B$ meson oscillations.
If the ATLAS or CMS vertexing systems could with sufficiently high efficiency measure the
displacement of the secondary vertex which produces the $\mu$, this would allow separation of samples
in which the $B$ meson has had time to oscillate (i.e. samples with a large displacement) 
from samples in which the decay time is short compared to the oscillation period.  These samples would
have different dilution factors and could be weighted differently to improve sensitivity.  
While we have not investigated such advanced techniques in our analysis, we present in \Fig{fig:tdl} a comparison of the transverse decay length 
for different sources of the muon, illustrating this effect.

Finally, we comment that while our analysis was designed with one of the LHC's all-purpose detectors in mind (i.e. ATLAS and CMS), 
it is worthwhile to consider the reach of LHCb as we are interested in a precision measurement 
of $b$-jets in the forward region.  The main distinguishing feature of LHCb is its precision tracker and vertexing system, which allows for a precise
reconstruction of hadron level decays.  If this could be used to probe the decays of the $b$-hadrons then it could allow for a substantial reduction in the 
wrong-sign $\mu$ backgrounds and may open up other channels for use in signing the $b$.  However, such a measurement would be challenging as 
the rates for $b\bar b$ production become quite small once one restricts both $b$s to lie in the forward region.  At parton level 
we find the $b\bar b$ rate to be $\sim 0.5~{\rm pb}$, yielding an asymmetry of $\sim3\%$ when requiring only $p_{T}(b/\bar b)>150~{\rm GeV}$ and $2<y(b/\bar b)<5$ (the rapidity range for LHCb), with the rate 
dropping precipitously as cuts on $m_{jj}$ are further applied.  Further challenges may also come from employing LHCb to study high-$p_T$ jets that we require, as the detector was primarily designed to 
study softer objects in a relatively clean environment.

The situation changes somewhat at a 14 TeV LHC, where the
parton level rate for $b\bar b$ production subject to the above cuts
rises to $12~{\rm pb}$, yielding an asymmetry of $\sim 5\%$ before accounting 
for other sources of background (i.e. gluon splitting, flavor excitation, and $b$-fakes).  Here
LHCb might be able to measure the asymmetry in $b$-production, although to properly evaluate its 
potential one would need to perform a more detailed study of its capabilities than we would feel comfortable making.  We
therefore feel that although it appears that such a measurement would be quite difficult, a more detailed study of LHCb's reach in this channel is probably warranted.

 \section{Conclusions}

The CDF and D0 collaborations have both observed an anomalously large asymmetry in $t\bar t$ production. 
While some discrepancies between the two experiments remain to be resolved,  the evidence for a large asymmetry
 seems robust, 
and if the excess is due to SM effects they must be quite subtle.  Many beyond the SM explanations have been put forward to explain the asymmetry, 
offering various treatments of the many potential  quark couplings to new-physics.  Previously, \Refs{Strassler:2011vr,Bai:2011ed} proposed that Tevatron data at CDF and D0 could be used to probe these interactions for bottom and charm quarks.  Here we have argued that, through a forward-central asymmetry,
 the CMS and ATLAS experiments at the LHC are sensitive in the immediate future to whether new-physics interactions generating the asymmetry in $t\bar t$ production also affect the bottom quark.

Our results indicate that, with around $10~{\rm
 fb}^{-1}$ of $7~{\rm TeV}$ LHC data,  the general purpose LHC detectors can probe
such new interactions with a sensitivity greater than $2\sigma$.
While less sensitive than a Tevatron search with the same amount of data,
and while insufficient to
 discover new physics, such a measurement would still provide useful
 model-building guidance.  However, whether this is feasible depends
 crucially upon whether the selection cuts required for the
 measurement are compatible with the trigger menu for the corresponding integrated luminosity.  Given the importance of determining whether there are unexpected asymmetries affecting bottom quark production, we hope
 that the ATLAS and CMS experiments will investigate this issue carefully,
and consider adjusting trigger thresholds if adjustments are indeed necessary.
\vspace{0.5 cm}


\acknowledgments{
We would like to thank Gustaaf Brooijmans, Valerie Halyo, David E. Kaplan, Sal Rappoccio, Matt Schwartz, and Sheldon Stone for helpful discussions.

D. Kahawala is supported by the General Sir John Monash Award.  D. Krohn is supported by a Simons postdoctoral fellowship and by an LHC-TI travel grant. M.J.S. is supported by NSF grant PHY-0904069 and by DOE grant DE-FG02-96ER40959}

\label{sec:conclude}
\appendix

\bibliography{tafb} 
\bibliographystyle{apsrev4-1}

\end{document}